# Quantum Cybernetics: A New Perspective for Nelson's Stochastic Theory, Nonlocality, and the Klein-Gordon Equation

Gerhard Grössing,
*Austrian Institute for Nonlinear Studies*,
Parkgasse 9, A 1030 Vienna, Austria
e-mail: ains@teleweb.at

**Abstract**

The Klein-Gordon equation is shown to be equivalent to coupled partial differential equations for a sub-quantum Brownian movement of a "particle", which is both passively affected by, and actively affecting, a diffusion process of its generally nonlocal environment. This indicates circularly causal, or "cybernetic", relationships between "particles" and their surroundings.
Moreover, in the relativistic domain, the original stochastic theory of Nelson is shown to hold as a limiting case only, i.e., for a vanishing quantum potential.

1. **Introduction**

The only existing derivation of the Schrödinger equation, i.e., Nelson's stochastic theory [1,2], is based on the assumption of the existence of some sub-quantum medium which would render "particle" trajectories as subject to some Brownian-type motion. (See also the paper by I. Fenyes [3] on an earlier variant of such an approach.) The introduction of "particle" trajectories is due to the de Broglie-Bohm interpretation (BBI) of quantum theory, and also Nelson's approach is related to this causal interpretation.

In fact, the two basic real-valued equations of the BBI, obtained after decomposition of the complex-valued Schrödinger equation, also is at the heart of Nelson's theory. However, by introducing a new type of differential calculus, Nelson attempted to *derive* these two real-valued equations on assuming that a "particle's" Brownian motion can be considered as a Markov process whose basic equations could be formulated with the aid of his differential calculus. More recently, the latter has been criticised by Bohm and Hiley [4], and, in fact, it is hard to see how Nelson's complete time symmetry for the stochastic dynamics, or the mathematical arbitrariness of his definition of acceleration, could be justified physically.

However, there remains the following very intriguing fact, which can be obtained in two steps: First, one decomposes the Schrödinger equation,

$$i\hbar \frac{\partial \Psi}{\partial t} = \left( -\frac{\hbar^2}{2m} \nabla^2 + V \right) \Psi , \qquad (1.1)$$

in the BBI manner into a) the conservation equation of the probability current,

$$\frac{\partial P}{\partial t} + \nabla \cdot (\mathbf{v}P) = 0, \qquad (1.2)$$

where

$$\mathbf{v} = \frac{\nabla S}{m},$$

$$P = R^2 = \Psi^*\Psi,$$

with

$$\Psi = Re^{iS/\hbar},$$

and b) into the Hamilton-Jacobi-Bohm equation,

$$\frac{\partial S}{\partial t} + \frac{(\nabla S)^2}{2m} + V + Q = 0, \qquad (1.3)$$

where $V$ is an external classical potential, and the additional "quantum potential" $Q$ is given by

$$Q = -\frac{\hbar^2}{2m}\frac{\nabla^2 R}{R}.$$

The second step then consists in taking the gradient of the two equations, (1.2) and (1.3), respectively. One obtains

$$\frac{\partial \mathbf{u}}{\partial t} = -D\nabla^2 \mathbf{v} - \nabla(\mathbf{u}\cdot\mathbf{v}) \qquad (1.4)$$

and

$$\frac{\partial \mathbf{v}}{\partial t} = \mathbf{a} - (\mathbf{v}\cdot\nabla)\mathbf{v} + (\mathbf{u}\cdot\nabla)\mathbf{u} + D\nabla^2 \mathbf{u}, \qquad (1.5)$$

with the "diffusion coefficient"

$$D := \frac{\hbar}{2m},$$

the "osmotic velocity"

$$\mathbf{u} = \frac{\hbar}{m}\frac{\nabla R}{R} = D\nabla \ln P,$$

and the classical acceleration

$$\mathbf{a} = -\frac{\nabla V}{m}.$$

Equations (1.4) and (1.5), in turn, and together with (1.2), represent nothing but the equations for Brownian motion of a particle suspended in some "medium".

More generally [5], one could also start with a Fokker-Planck equation for the sub-quantum medium,

$$\frac{\partial P}{\partial t} + \nabla \cdot (\mathbf{b}P - \nu\nabla P) = 0. \tag{1.6}$$

Here, one assumes a diffusion process for the "particle", such that

$$d\mathbf{x} = \mathbf{b}dt + \sqrt{\alpha}d\boldsymbol{\omega}, \tag{1.7}$$

where

$$\mathbf{b} = \frac{\hbar}{m}\nabla S + \alpha\frac{\hbar}{2m}\frac{\nabla |\Psi|^2}{|\Psi|^2}$$

and $d\boldsymbol{\omega}$ represents a Wiener process, i.e., $\overline{d\boldsymbol{\omega}} = 0$ and $\overline{d\boldsymbol{\omega}^2} = \frac{\hbar}{m}$.

With $\mathbf{b}$ being the drift velocity and $\nu := \alpha\frac{\hbar}{2m} = \alpha D$ the diffusion coefficient, the current velocity is given by

$$\mathbf{v} = \frac{\hbar}{m}\nabla S, \tag{1.8}$$

whereas the osmotic velocity

$$\mathbf{u} = \nu \frac{\nabla |\Psi|^2}{|\Psi|^2}. \tag{1.9}$$

Then, by substituting $P = |\Psi|^2$, equation (1.6) reduces to the usual equation for current conservation, equation (1.3):

$$\frac{\partial |\Psi|^2}{\partial t} + \nabla \cdot \left( \frac{\hbar}{m} (\nabla S) |\Psi|^2 \right) = 0.$$

This leaves $\rho := |\Psi|^2$ as an equilibrium solution which is always preserved by the BBI dynamics, and which represents the situation that osmotic current $\mathbf{u}P$ and diffusion current $\nu \nabla P$ always balance each other exactly:

$$\mathbf{u}P = \nu \nabla P. \tag{1.10}$$

(Remark: As noted in [5], from equation (1.7) one obtains the usual BBI for $\alpha = 0$, and Nelson's original theory for $\alpha = 1$.)

Still, a basic problem with the Nelsonian approach remains. As Nelson himself concedes [1,6], his theory is purely local and thus hardly compatible with the more recent experimental evidence of quantum mechanical nonlocality.

Moreover, a generally accepted relativistic formulation of Nelson's theory has not yet been achieved. In this paper, I shall thus try a new approach to the Klein-Gordon equation from a Nelsonian perspective. In doing so, the basic concepts of *quantum cybernetics* [7] will be implemented, which turn out to provide a useful extension of the usual BBI interpretation with regard to Nelson's stochastic theory. In particular, the Klein-Gordon equation will be seen to be equivalent to equations for a sub-quantum Brownian movement of a "particle" which is both passively affected by, and actively affecting, a diffusion process of its generally nonlocal environment.

## 2. Vacuum energy and the causal formulation of the Klein-Gordon theory

Relativistic versions of the BBI had for some time in the past been considered as unobtainable without additional assumptions. However, in recent years, both Dirac and Klein-Gordon equations have found satisfactory causal formulations. As for the Dirac theory, a solution was presented by Holland [8,9], whereas the Klein-Gordon case is treated by Horton *et al.* [10]. However, the latter reference has to introduce the causal description of time-like flows for the Klein-Gordon equation in an Einstein-Riemann space. That is, whereas a causal Dirac theory can be formulated in Minkowski space, this apparently does not hold for the Klein-Gordon theory. (Basically, the probability current can assume negative values of its zeroth component and is not generally timelike.)

There exists a stochastic derivation of the Klein-Gordon equation in Minkowski space by Lehr and Park [11], but this derivation is based on a) a Nelsonian differential calculus (which, as mentioned above, has no immediate physical interpretation) and b) two additional assumptions (axioms) which definitely go beyond the broadly accepted theory (i.e., (i) discretization of time, and (ii) a very specific *ad hoc* model for particle-thermostat interactions). Moreover, in substituting the Markovian by the more encompassing Bernstein processes, Zambrini [12] and Serva [13] manage to overcome some, but not all, of the limitations of Nelson's original theory. (As Bernstein processes can be considered, and are in fact sometimes referred to, as "reciprocal", even a link to circular causal modelling as discussed in this paper may be feasible.)

However, as was already indicated in reference [7], pp. 67 f, one can envisage a causal Klein-Gordon theory in Minkowski space (thereby achieving a common level with the Dirac theory), if one introduces just one additional assumption. In fact, said assumption becomes more and more likely to hold true on the ground of empirical evidence from cosmology. There, it has become ever more likely throughout recent years that the quantum vacuum in some way has to play a major role in the evolution of the universe. In other words, one generally accepts the necessity to take the vacuum energy seriously by either re-introducing Einstein's $\Lambda$- term in the field equations of gravitation, or by introducing a more dynamic variant thereof, usually called "quintessence". [14,15]

Thus, we are in today's cosmology confronted with a re-introduction, although with different characteristics, of what earlier has been called the "aether". And, in fact, it is also the assumption of an "aether" which characterizes the BBI. Actually, it is very straightforward to see how the vacuum energy can enter a relativistic version of the BBI, and this is what shall be shown here in the context of the Klein-Gordon theory.

We start with the usual initial procedure of the BBI approach, i.e., by writing down the solution

$$\Psi(x,t) = R(x,t) e^{-iS/\hbar} \tag{2.1}$$

of the Klein-Gordon equation

$$\left(\Box + \frac{m^2 c^2}{\hbar^2}\right)\Psi = 0. \tag{2.2}$$

Inserting $(2.1)$ into $(2.2)$, and dividing by $R$, one obtains:

$$\frac{\Box R}{R} + \frac{i}{\hbar}\frac{\partial_\mu R}{R}\partial^\mu S + \frac{i}{\hbar}\frac{\partial^\mu R}{R}\partial_\mu S - \frac{1}{\hbar^2}\partial_\mu S \partial^\mu S + \frac{i}{\hbar}\Box S + \frac{m^2 c^2}{\hbar^2} = 0. \qquad (2.3)$$

Now one solves equation (2.3) by separating the real-valued and the complex-valued parts, respectively, and by setting both of them equal to zero, so as to obtain the well-known real-valued probability current and Hamilton-Jacobi-Bohm equations, respectively. In this way, from equation (2.3) one obtains

$$2\frac{\partial^\mu R}{R}\partial_\mu S + \Box S = 0, \qquad (2.4)$$

and thus, with $P = R^2$ and $J_\mu := P\partial_\mu S$,

$$\partial^\mu J_\mu = \partial^\mu P \partial_\mu S + P \Box S = 0, \qquad (2.5)$$

i. e., the conservation of the probability current $J_\mu$, as well as the Hamilton-Jacobi-Bohm equations

$$\partial_\mu S \partial^\mu S =: M^2 c^2 = m^2 c^2 + \hbar^2 \frac{\Box R}{R}. \qquad (2.6)$$

Now we want to add an extra phase term proportional to $S^0$ in the exponent of equation (2.1), such as to explicitly take into account the vacuum energy, i. e.,

$$S^0 = E_{vac} t. \qquad (2.7)$$

That is, we now use as a solution of the Klein-Gordon equation the wave function

$$\Psi(x,t) = R(x,t) e^{-i(S+S^0)/\hbar}. \qquad (2.8)$$

However, with our extra term, one can now show the viability of a causal Klein-Gordon theory in Minkowski space. As already mentioned, our additional input consists of the assumption that said constant is determined by some "vacuum background" energy which is supposed to fill all space.

In fact, instead of equation (2.5), we now obtain an equation which expresses the conservation of the ordinary probability current $J_\mu$ plus some "hidden" probability current $J_\mu^0 := P\partial_\mu S^0$:

$$\partial^\mu J_\mu = \partial^\mu P \partial_\mu S + P\Box S = -\partial^\mu P \partial_\mu S^0 - P\Box S^0 = -\partial^\mu J_\mu^0. \tag{2.9}$$

In other words, one now has a conservation law for the total probability current

$$\partial^\mu \left( J_\mu + J_\mu^0 \right) = 0. \tag{2.10}$$

Moreover, the new term $S^0$ now also modifies equation (2.6). Here it leads to a term due to an external potential in the energy balance for the expression of the total system's "variable mass" $M$ (or "variable energy", respectively) such that now

$$M^2 c^2 \to \left( mc + \frac{E_{vac}}{c} \right)^2 + \hbar^2 \frac{\Box R}{R}. \tag{2.11}$$

We thereby introduce a ("hidden") diffusion current $J_\mu^0$ due to the assumed stochastic aether dynamics on one hand, and a corresponding all-pervading vacuum energy $E_{vac}$ on the other hand. Note that the unknown quantity $E_{vac}$ can actually be very large, such that the variable mass/energy in (2.11) is always positive definite and always produces timelike trajectories, i. e., also the zero-component of the total current $J_{tot}^0 = P \frac{\partial}{\partial t} \left( S + S^0 \right) \geq 0$. Extension to many-particle systems is straightforward.

In other words, this essentially constitutes a working Klein-Gordon theory in Minkowski space.

However, in this paper I will not follow this line of arguments any further, and use the standard Klein-Gordon equation instead, because the relation to Nelson's theory can be seen already this way, and also because comparison with existing literature is made easier. Basically, my intention in this chapter was to show that it makes perfect sense to speak of a causal Klein-Gordon theory in Minkowski space.

## 3. Nelson's stochastic theory as a limiting case of the causal Klein-Gordon theory

Let us now tackle the question of the relationship between a causal Klein-Gordon theory and Nelson's stochastic theory. First, we define

$$S := Et - \mathbf{p}\mathbf{x} = \hbar k_\mu x^\mu, \tag{3.1}$$

such that

$$v^\mu = \frac{\partial^\mu S}{M}, \tag{3.2}$$

and, assuming some non-vanishing external potential $V$ (which could include the vacuum energy term of equation (2.11)), with $M$ given by equation (2.6) as

$$M = \sqrt{\left(m + \frac{V}{c^2}\right)^2 + \frac{\hbar^2}{c^2}\frac{\Box R}{R}}. \tag{3.3}$$

We thus obtain the following form of the two real-valued analogues of the Klein-Gordon equation. The conservation of the probability current reads as

$$\partial^\mu J_\mu = \partial^\mu P \partial_\mu S + P \Box S = 0, \tag{3.4}$$

and the Hamilton-Jacobi-Bohm equation is given by

$$\partial_\mu S \partial^\mu S = M^2 v_\mu v^\mu = M^2 c^2 = \left(mc + \frac{V}{c}\right)^2 + Q, \tag{3.5}$$

with the "quantum potential"

$$Q = \hbar^2 \frac{\Box R}{R}. \tag{3.6}$$

Then, by generalizing the non-relativistic expression of equation (1.10) into the relativistic domain, we define the "osmotic velocity"

$$u^\mu := D \frac{\partial^\mu P}{P}, \tag{3.7}$$

where we may choose the conventional expression for the diffusion constant, i.e.,

$$D = \frac{\hbar}{2m}. \tag{3.8}$$

(Note that, apart from the sign and different arguments for its introduction, an expression of the form (3.7) has already been put forward in reference [7], page 69, to account for a space-like velocity $u^\mu := \frac{1}{\gamma}(c, \mathbf{u})$ orthogonal to a "particle's" velocity $v_\mu := \gamma(c, \mathbf{v})$, where $\gamma = 1/\sqrt{1 - v^2/c^2}$.)

With our definition of the osmotic velocity, we can now rewrite equations (3.4) and (3.5), respectively, as

$$\partial_\mu v^\mu = -\frac{1}{D} u_\mu v^\mu - \frac{\partial_\mu M}{M} v^\mu \tag{3.9}$$

and

$$M^2 c^2 = \left(mc + \frac{V}{c}\right)^2 + m\hbar \partial_\mu u^\mu + m^2 u_\mu u^\mu. \tag{3.10}$$

(Note that for timelike velocities $u^\mu = \gamma(c, \mathbf{u})$, the expression for the quantum potential, given by the last two terms on the r.h.s. of equation (3.10), reads as

$$Q = m^2 c^2 + m\hbar \partial_\mu u^\mu,$$

which means that the dynamics is essentially governed by the four-gradient of the osmotic velocity.)

In the sense of Lehr and Park [11], but now in explicit dependence of the osmotic velocity, equations (3.9) and (3.10) are the *two basic equations of relativistic stochastic mechanics*. Note that equation (3.9) in particular provides an elegant and compact description of the role of the osmotic velocity relative to the "particle's" velocity. Furthermore, it also provides a clear expression as to the bridge between the quantum and classical regimes, respectively. As was shown in reference [7], in (classical) special relativity it always holds for unaccelerated, conservative systems (i.e., where $\partial_\mu p^\mu := \partial_\mu (M \text{v}^\mu) = 0$) that if $M = const.$, $\partial_\mu \text{v}^\mu = 0,$ and thus, according to equation (3.9),

$$\text{u}_\mu \text{v}^\mu = c^2 - \mathbf{uv} = 0. \tag{3.11}$$

In other words, current and osmotic velocities are then always orthogonal to each other. This, in turn, is equivalent to a) the current conservation

$$\frac{\partial}{\partial t} P = -\mathbf{v} \cdot \nabla P, \tag{3.12}$$

and b) the expression for the propagation of "phase waves" ($S = const.$) with velocity $\mathbf{u}$, i.e.,

$$\frac{\partial}{\partial t} S = -\mathbf{u} \cdot \nabla S. \tag{3.13}$$

(To verify, multiplication of (3.12) with (3.13) provides $\partial_\mu P \partial^\mu S = 0,$ and thus (3.11).)

In still other words, classically the osmotic velocity corresponds to the speed of Rindler's "waves of simultaneity" [7]. In the quantum domain, however, current conservation (3.4) can hold even if $M = M(x,t) \neq const.$, and one has to use equation (3.9) instead of equation (3.11). That is, while $\text{u}_\mu$ and $\text{v}^\mu$ can still be orthogonal,

$$\partial_\mu v^\mu = -\frac{\partial_\mu M}{M} v^\mu \neq 0.$$

Moreover, for energetically open systems (where $\partial_\mu p^\mu \neq 0$), equation (3.9) provides

$$\frac{1}{D} u_\mu v^\mu = -\left(\frac{\partial_\mu M}{M} + \partial_\mu\right) v^\mu \neq 0, \qquad (3.14)$$

i.e., $u_\mu$ and $v^\mu$ will in general not be orthogonal. In fact, in this case the diffusion velocity can be timelike, as will be shown below.

Note also that one can rewrite equation (3.9) as

$$\partial_\mu v^\mu = -\frac{1}{D}\left(u_\mu + u_\mu^M\right) v^\mu, \qquad (3.15)$$

where

$$u_\mu^M := D \frac{\partial_\mu M}{M}. \qquad (3.16)$$

Now let us continue with the same procedure as already discussed in chapter 1 for the nonrelativistic case, i.e., now taking the gradients of the relativistic equations (3.9) and (3.10). In a first step we restrict ourselves to constant "particle" masses/energies, i.e., $m = \text{constant}$, spacelike velocities $u$ larger than, and timelike velocities $v$ smaller than $c$ (i.e., such that equation (3.11) holds). We also note that for spacelike velocities $u_\mu$, which are associated with "particle" velocities $v_\mu$, it holds [7] that $u_\mu u^\mu = -(\mathbf{v} - \mathbf{u})^2$. Then, using $\partial S/\partial t = Mc^2$, $\nabla S = -M\mathbf{v}$, and equations (3.2) and (3.7), we obtain (for "particles" with velocities $\frac{dx}{dt} = v$ and "antiparticles" with velocities $\frac{dx}{d(-t)} =: -v$):

$$\frac{\partial \mathbf{u}}{\partial t} = -D\nabla^2 \mathbf{v} - \nabla(\mathbf{v} \cdot \mathbf{u}) \qquad (3.17)$$

and

$$\frac{\partial \mathbf{v}}{\partial t} = \mathbf{a} \pm (\mathbf{v} \cdot \nabla)\mathbf{v} + (\mathbf{u} \cdot \nabla)\mathbf{u} + D\nabla^2 \mathbf{u}, \qquad (3.18)$$

*provided* that $M = m$. Thus, we see that we obtain exactly the same equations as Nelson, viz., equations (1.4) and (1.5), for spacelike velocities $u_\mu$, but only under the specific requirement that $M = m$, which means, in other words, that in the absence of a classical potential $V$, the quantum potential $Q$ must vanish, too. (In fact, even if $V \neq 0$, solutions for a non-vanishing $Q$ would require a permanent matching with the quantities $m$ and $V$, which would be highly unphysical.) Note, however, that whereas the classical acceleration is given by $\mathbf{a} = -V\nabla V/M^2 c^2$, the complete acceleration is

$$\frac{\partial \mathbf{v}}{\partial t} = -\frac{\nabla M}{M} c^2. \qquad (3.19)$$

Thus, when $M = const.$, the r.h.s. of equation (3.18) vanishes, so that there will be no acceleration. This holds particularly for situations with vanishing quantum potential.

If, on the other hand, both $v_\mu$ and $u_\mu$ are timelike, i.e., $v_\mu v^\mu = u_\mu u^\mu = c^2$, we obtain again equation (3.17), but instead of (3.18) we now have, again for $M = m$, that

$$\frac{\partial \mathbf{v}}{\partial t} = \mathbf{a} + D\nabla^2 \mathbf{u}. \qquad (3.20)$$

Comparing with equations (1.4) and (1.5), this means that we again obtain the equations for Brownian motion, with the condition that

$$\mp (\mathbf{v} \cdot \nabla)\mathbf{v} = (\mathbf{u} \cdot \nabla)\mathbf{u}. \qquad (3.21)$$

## 4. Circular causality in the sub-quantum domain: Quantum Cybernetics

Now we allow for the general situation that $M \neq const.$, and for a variable rest energy, too, i.e., also $m \neq const.$, as it exists for a "particle in a box" with moving walls, for example [7]. Taking the gradients of the relativistic equations (3.9) and (3.10) now provides the general solutions

$$\frac{\partial \mathbf{u}}{\partial t} = -D\nabla^2 \mathbf{v} - \nabla(\mathbf{v} \cdot \mathbf{u}) + D(\nabla \cdot \mathbf{v})\left[\frac{\nabla m}{m} - \frac{\nabla M}{M}\right]$$
$$+ D\frac{\nabla m}{m}\left[\frac{\dot{M}}{M} + \mathbf{v} \cdot \frac{\nabla M}{M}\right] - D\left[\nabla\left(\frac{\dot{M}}{M}\right) - (\mathbf{v} \cdot \nabla)\left(\frac{\nabla M}{M}\right)\right], \quad (4.1)$$

and, for spacelike velocities $\mathbf{u}$,

$$\frac{\partial \mathbf{v}}{\partial t} = \mathbf{a} + D\frac{m^2}{M^2}\left[\nabla^2 \mathbf{u} + \mathbf{u}\frac{\nabla^2 m}{m} + 2\nabla \mathbf{u}\frac{\nabla m}{m} - \frac{\nabla \dot{m}}{m}\right]$$
$$+ \frac{m^2}{M^2}\left[(\mathbf{u} \cdot \nabla)\mathbf{u} \pm (\mathbf{v} \cdot \nabla)\mathbf{v}\right] - \frac{m\nabla m}{M^2}(2c^2 - \mathbf{u}^2 - \mathbf{v}^2), \quad (4.2)$$

whereas for timelike $\mathbf{u}$ one obtains

$$\frac{\partial \mathbf{v}}{\partial t} = \mathbf{a} + D\frac{m^2}{M^2}\left[\nabla^2 \mathbf{u} + \mathbf{u}\frac{\nabla^2 m}{m} + 2\nabla \mathbf{u}\frac{\nabla m}{m} - \frac{\nabla \dot{m}}{m}\right] - \frac{m\nabla m}{M^2}c^2. \quad (4.3)$$

(For $M = m = const.$, equations (4.1), (4.2), and (4.3) reduce to the equations (3.17), (3.18), and (3.20), respectively.)

Note that for equation (4.1), if $m \neq const.$, one can define a new velocity

$$\mathbf{u}_D := D\frac{\nabla m}{m} = -\nabla D, \tag{4.4}$$

such that (4.1) becomes, for $M = const.$,

$$\frac{\partial \mathbf{u}}{\partial t} = -D\nabla^2 \mathbf{v} - \nabla(\mathbf{v} \cdot \mathbf{u}) + \mathbf{u}_D(\nabla \cdot \mathbf{v}). \tag{4.5}$$

This is particularly interesting for the case of a particle in a box of length $L$ with ("practically") infinitely high walls, where one wall of the box may be moved. In this case, we still have the energy eigenvalues for a particle of mass $m = \hbar\omega/c^2$ as

$$E_n = \hbar\omega_n = \frac{\hbar^2 n^2 \pi^2}{2mL^2}, \tag{4.6}$$

such that

$$\frac{\nabla m}{m} = -\frac{\nabla L}{L}. \tag{4.7}$$

Thus, the velocity $\mathbf{u}_D$ of equation (4.4), which is nothing but the gradient of the diffusion constant, is essentially given by the expression of equation (4.7). In other words, the variation of the length $L$ of our box corresponds to a variation of the diffusion constant $D$, thereby introducing the velocity $\mathbf{u}_D$.

It was implied above that in conservative systems the osmotic velocity $\mathbf{u}$ can be infinite. However, when opening our system by moving one wall of our box, the probability distribution $P$ for each eigenvalue $n$,

$$P = \frac{2}{L}\sin^2\left(\frac{n\pi x}{L}\right) \tag{4.8}$$

provides

$$\frac{\nabla P}{P} = -\frac{\nabla L}{L} + 2\cot\left(\frac{n\pi x}{L}\right)\left[\frac{n\pi}{L} - \frac{n\pi x}{L}\frac{\nabla L}{L}\right], \tag{4.9}$$

such that for $\nabla L = 0$, i.e., for the conservative system, we get with $\frac{n\pi}{L} = k = \frac{m\mathbf{v}}{\hbar}$:

$$\mathbf{u} = \frac{\hbar}{2m}\frac{\nabla P}{P} = \mathbf{v}\cot\left(\frac{n\pi x}{L}\right). \quad (4.10)$$

In this case, $\mathbf{u}$ still tends towards infinity around the nodes of the probability distribution (4.8). However, further off the nodes, and particularly for the energetically open scenario when $\nabla L \simeq L/x \neq 0$, such that the second term on the r.h.s. of equation (4.9) vanishes, we obtain

$$\mathbf{u} = -\frac{\hbar}{2m}\frac{\nabla L}{L} = D\frac{\nabla m}{m} = -\nabla D = \mathbf{u}_D \ll c. \quad (4.11)$$

In short, while the osmotic velocity can be spacelike for conservative systems, it can become timelike for open systems.

## 5. Conclusions and Outlook

How can one picture the underlying physics of the results just presented? First note that classically diffusion waves involve coherent, always driven, oscillations of diffusing energy. We have already discussed above, what may drive oscillations at the sub-quantum level, i.e., the quantum vacuum energy, or some "aether dynamics", respectively. In this picture, a "particle" of frequency $\omega = mc^2/\hbar$ oscillates in phase with the whole surrounding medium. In fact, with our definition of the osmotic velocity of equation (3.7), we can also define a balancing diffusion current

$$\mathbf{J}_D = P\mathbf{u} = PD\frac{\nabla P}{P} = D\nabla P. \tag{5.1}$$

This is Fick's first law of diffusion. Inserting the corresponding four-current $J_D^\mu = P\mathrm{u}^\mu$ into the continuity equation $\partial_\mu J^\mu = 0$ then provides Fick's second law of diffusion, i.e.,

$$\frac{\partial}{\partial t}P = D\nabla^2 P. \tag{5.2}$$

Due to the linear law (5.1), rather than the square laws of the usual wave propagations, diffusion waves at interfaces obey an accumulation-depletion law, rather than the usual reflection-refraction law. Moreover, it is well known that classical diffusion waves can exhibit infinite speeds, albeit with very small amplitude, at remote locations away from the source. [16] In this regard, experiments have proven the existence of sudden perturbations over entire domains. [17, 18] The diffusion waves are only characterized by spatially correlated phase lags. And, as Andreas Mandelis explains [17], "even stranger properties emerge. Because propagation is instantaneous, the equations yield no travelling waves, no wavefronts, and no phase velocity. Rather, the entire domain 'breathes' in phase with the oscillating source." All these features, therefore, seem particularly amenable to describe the sub-quantum

"aether dynamics" given by the balance between osmotic and diffusion currents as discussed in section 1.

Finally, we note that whenever in the equations presented here the expression $\nabla m/m$ is not equal to zero, we can conclude that there exists a *dynamical* effect of the variation of a "particle's" central frequency $\omega = mc^2/\hbar$ such that the characteristic feature of quantum cybernetics is shown to hold: Not only do the wave configurations of the surrounding medium influence the path of a "particle", but also the reverse is true, i.e., a change in the "particle's" central frequency is transmitted with osmotic velocity into the surrounding medium. Thus, "particles" and waves relate to each other on an equal footing. [7]

As an example, consider some uniform mass $m$ of exponentially (but not necessarily radioactively) "decaying" material, consisting of such a large number $N$ of particles that mathematically the use of a mass gradient is feasible. Then, as $m \simeq m_0 e^{-\lambda t}$, we obtain

$$\frac{\nabla m}{m} = -\frac{\lambda}{\mathrm{v}_d}, \qquad (5.3)$$

where $\mathrm{v}_d$ is some decay velocity. This provides some extra terms in the expressions for current and osmotic velocities, respectively. To give an example, equation (4.5) now reads

$$\frac{\partial \mathbf{u}}{\partial t} = -D\nabla^2 \mathbf{v} - \nabla(\mathbf{v}\cdot\mathbf{u}) - \frac{D\lambda}{\mathrm{v}_d}(\nabla\cdot\mathbf{v}). \qquad (5.4)$$

Effects of varying "particle" mass on the overall dynamics of the quantum system could thus eventually be observable in experiment. In any case, investigations along these lines should eventually pave the way for a physics beyond present-day quantum mechanics. One might well call it an "aether dynamics" then.